\documentclass[11pt]{article}	
\pdfoutput=1
\usepackage{jcappub}

\def\wisk#1{\ifmmode{#1}\else{$#1$}\fi}

\def\deg    {\wisk{^\circ}}

\def\aap {Astronomy and Astrophysics}
\def\apj {The Astrophysical Journal}
\def\apjs {The Astrophysical Journal Supplement Series}
\def\prd {Physical Review D}

\title{Calibration Method and Uncertainty for the
Primordial Inflation Explorer (PIXIE)}

\author[1]{A. Kogut}
\author[1,2]{and D.J. Fixsen}

\affiliation[1]{Code 665, 			
		Goddard Space Flight Center,	
		Greenbelt, MD 20771 USA }
\affiliation[2]{University of Maryland,		
		College Park, MD 20742 USA }

\emailAdd{Alan.J.Kogut@nasa.gov}
\emailAdd{Dale.J.Fixsen@nasa.gov}

\abstract{
The Primordial Inflation Explorer (PIXIE)
is an Explorer-class mission concept
to measure cosmological signals
from both linear polarization of the cosmic microwave background
and spectral distortions from a perfect blackbody.
The targeted measurement sensitivity is 2--4 orders of magnitude
below competing astrophysical foregrounds,
placing stringent requirements on instrument calibration.
An on-board blackbody calibrator
presents a polarizing Fourier transform spectrometer
with a known signal
to enable conversion of the sampled 
interference fringe patterns
from telemetry units to physical units.
We describe the instrumentation and operations
needed to calibrate PIXIE,
derive the expected uncertainty for the
intensity, polarization, and frequency scales,
and show the effect of calibration uncertainty
in the derived cosmological signals.
In-flight calibration is expected to be accurate
to a few parts in $10^6$
at frequencies dominated by the CMB,
and a few parts in $10^4$
at higher frequencies
dominated by the diffuse dust foreground.
}


\notoc

\begin{document}
\maketitle

\renewcommand{\bottomfraction}{0.9}
\renewcommand{\topfraction}{0.9}

\flushbottom

\section{Introduction}
The cosmic microwave background (CMB) provides a unique window
to the early universe.
Its blackbody spectrum points to a hot, dense phase 
in the early universe,
while spatial maps of small temperature perturbations
about the blackbody mean
provide detailed information
on the geometry, constituents, and evolution of the universe.

New measurements could provide additional insight.
Maps of CMB polarization trace a stochastic background of
gravitational radiation produced during an inflationary epoch
in the early universe,
testing physics at energies above $10^{15}$ GeV
while providing the first observational evidence 
for quantum gravity\cite{
lyth/riotto:1999,
krauss/wilczek:2014a,
krauss/wilczek:2014b}.
Small deviations from the monopole blackbody spectrum
(spectral distortions)
record energy transfers
between the evolving matter and radiation fields
to detail the thermal history of the universe\cite{
distortion_wp_2019}.

\begin{figure}[b]
\centerline{
\includegraphics[width=5.0in]{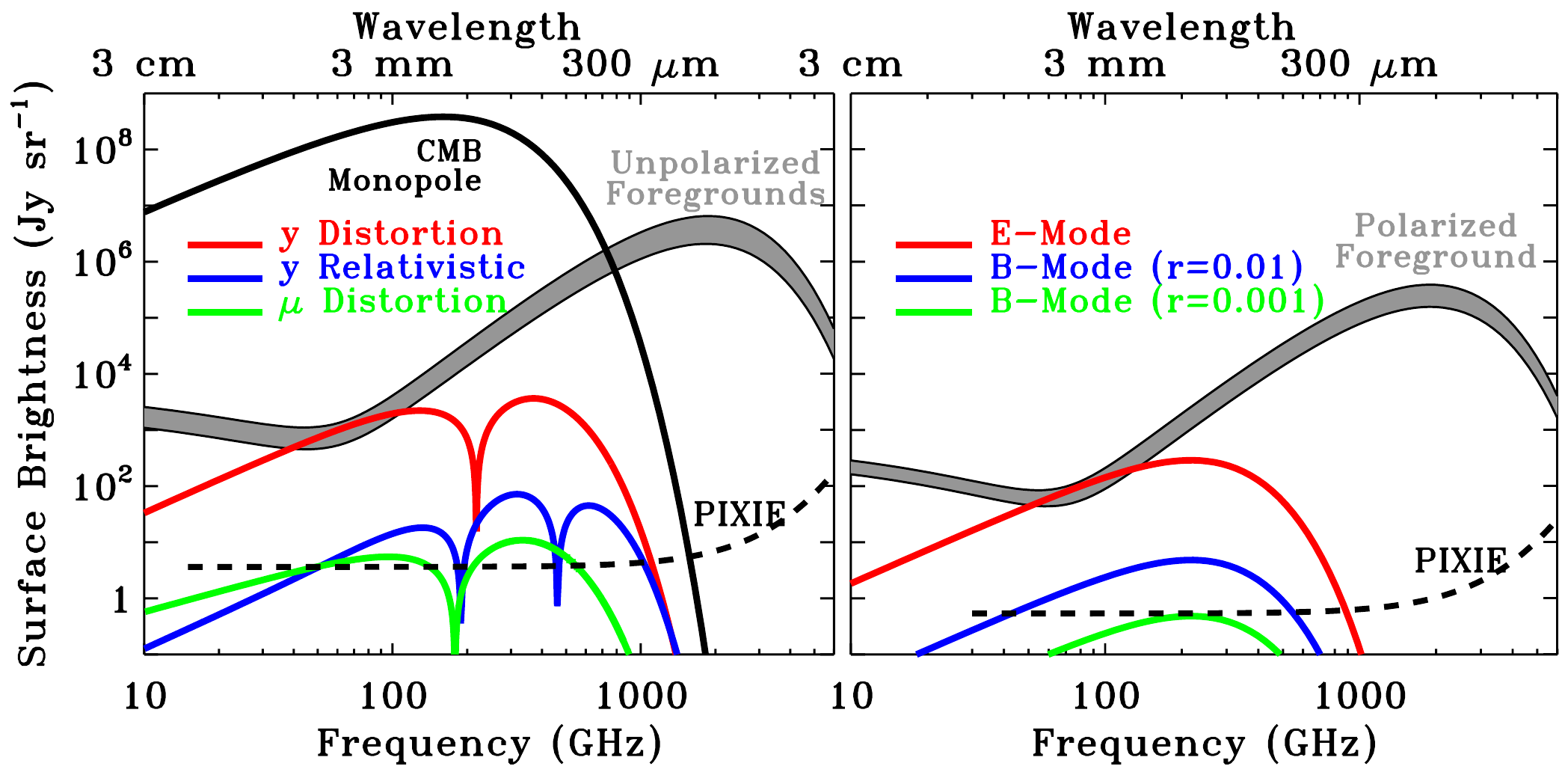}}
\caption{
Cosmological signals are faint compared to the
combined foreground emission.
(left) Unpolarized CMB spectral distortions
from the Compton y distortion caused by
scattering in groups and clusters,
the relativistic correction to the y distortion,
and the $\mu$ distortion
from dissipation of primordial density perturbations.
(right) Polarized signals from 
density perturbations (E-modes)
and inflation (B-modes).
The PIXIE noise level is shown
for 12 months cumulative integration time
in each observing mode.
}
\label{fg_figure}
\end{figure}

The cosmological signals
for both polarization and spectral distortions
are small compared to astrophysical foregrounds
(Figure \ref{fg_figure}).
CMB polarization can be decomposed
into a parity-even (E-mode) component
sourced by primordial density perturbations
and a parity-odd (B-mode) component
sourced by inflation.
For the simplest (single-field) inflation models,
the amplitude of the B-mode signal 
depends on the energy scale of inflation as
\vspace{-2mm}
\begin{equation}
E = 1.06 \times 10^{16} 
\left( 
\frac{r}{0.01} 
\right)^{1/4}
~{\rm GeV} 
\label{potential_eq}
\vspace{-2mm}
\end{equation}
where 
$r$ is the power ratio of gravitational waves to density fluctuations
\cite{lyth/riotto:1999}.
Current measurements set upper limits $r < 0.07$
\cite{bicep2xkeck_2016},
while
next-generation experiments look for sensitivity
$r \ll 10^{-3}$.
Spectral distortions
result from energy injection in the early universe
that drives the matter and radiation fields 
out of thermal equilibrium.
The simplest distortions result 
from Compton scattering of CMB photons from
relativistic electrons
and are characterized by the
Compton $y$ parameter for the optically-thin case
and 
the Bose-Einstein chemical potential $\mu$
for the optically thick case\cite{
zeldovich/sunyaev:1969,
sunyaev/zeldovich:1970}.
Current upper limits
$|y| < 15 \times 10^{-6}$
and
$| \mu| < 9 \times 10^{-5}$
correspond to fractional deviations
$\Delta I/I$ less than 50 parts per million 
in the CMB blackbody spectrum\cite{fixsen/etal:1996}.
Measurements with background-limited detectors
could improve these limits by several orders of magnitude,
opening a broad discovery space\cite{
distortion_wp_2019,
kogut_apc_2019}.

Distinguishing cosmological signals
from the competing foregrounds
based on their different frequency dependence
requires measurements over a broad frequency range.
At levels of a few nK in thermodynamic temperature
(1~Jy~sr$^{-1}$
or
10$^{-26}$ W m$^{-2}$ Hz$^{-1}$ sr$^{-1}$),
the targeted measurement sensitivity
for next-generation measurements
is 2--4 orders of magnitude or more
below the foreground amplitude,
placing stringent requirements on
instrument dynamic range and calibration.

Calibration presents an instrument with a known signal,
enabling the conversion of sampled data
from digitized telemetry units to physical units.
In-flight calibration of previous space-based CMB missions
has relied on observations of
on-board blackbody sources
\cite{firas_cal_1994},
astrophysical sources such as the Moon or planets
\cite{dmr_cal_1992, wmap_planets_2011, planck_hfi_cal_2014},
or the CMB dipole induced by the motion
of the spacecraft or Solar System
with respect to the CMB rest frame
\cite{dmr_cal_1992, wmap_cal_2003, planck_lfi_cal_2013, planck_hfi_cal_2014}.
Three aspects of the calibration are particularly important:
The \emph{absolute calibration}
which converts data from digitized telemetry units
to physical units,
the \emph{relative calibration}
or uncertainty in the calibration
between measurements in different frequency channels,
and the \emph{frequency calibration}
or extent to which the
frequency scale
and
passbands of individual frequency channels are known.

The Primordial Inflation Explorer (PIXIE)
is an Explorer-class mission concept to characterize 
both CMB polarization and spectral distortions
\citep{kogut/etal:2011}.
PIXIE represents an updated, fully symmetric version
of the seminal 
Far Infrared Absolute Spectrophotometer (FIRAS)
flown on the Cosmic Background Explorer\cite{boggess/etal:1992}.
Room-temperature versions\cite{pan/etal:2019}
validate the the optical design,
which we supplement using simulations 
and data from balloon-borne cryogenic instruments\cite{
arcade_cal_2006,
fixsen/etal:2011}.
This paper describes the instrumentation and techniques
to calibrate the PIXIE data,
derives the expected uncertainty
in the calibration,
and shows the effect of calibration uncertainty
on the extracted cosmological signals.

\section{The PIXIE Instrument}

\begin{figure}[b]
\centerline{
\includegraphics[width=5.0in]{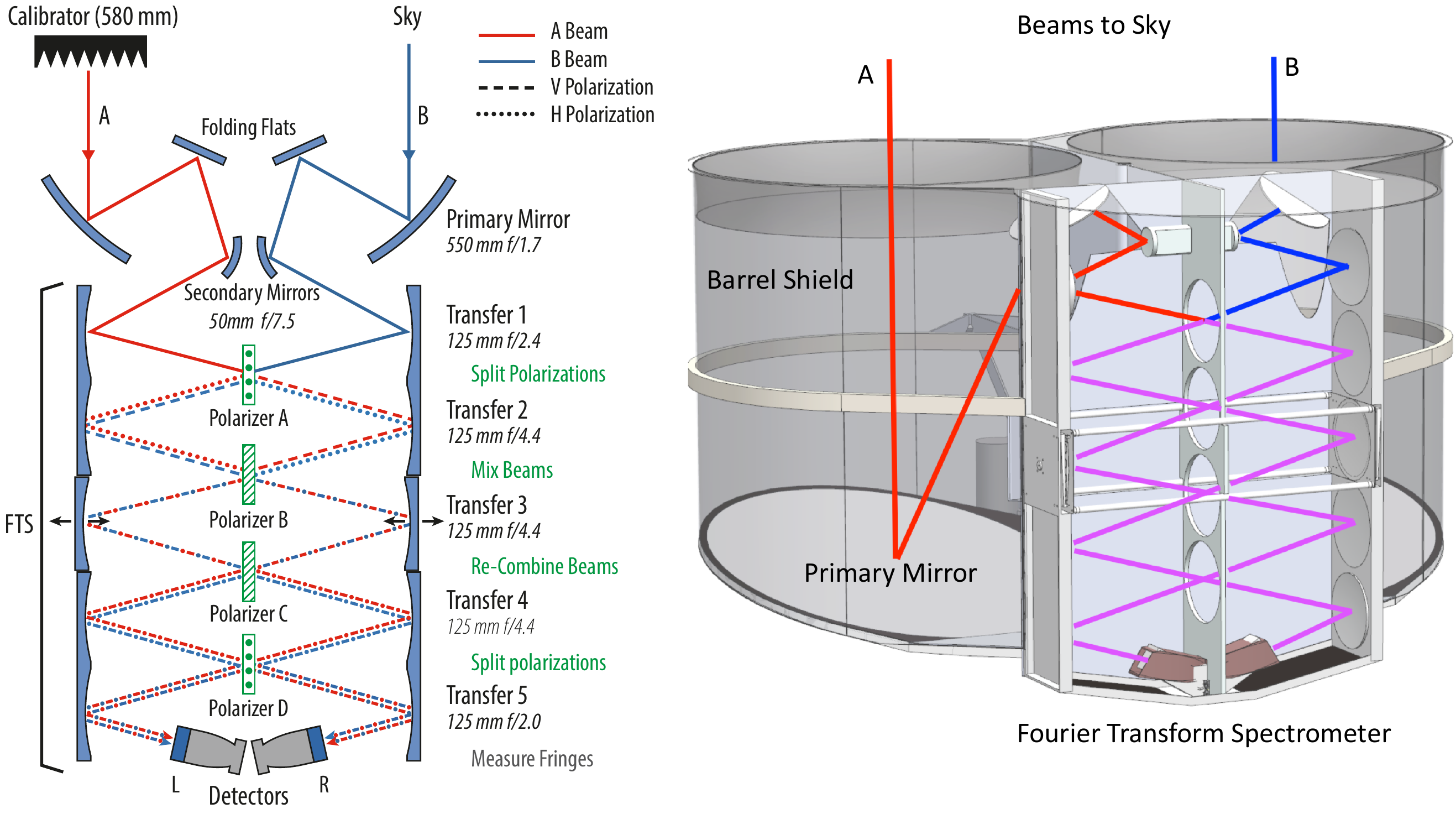}}
\caption{
Schematic rendering of the PIXIE instrument,
showing the proposed optical path (left)
and the beam-forming optics (right).
The signal path 
after the first polarizer
consists of linear combinations
of orthogonal polarization states (dashed vs dotted lines)
from the two input  ports
(red vs blue lines).
Linear combinations on the left side of the instrument
are orthogonal to those on the right side.
}
\label{pixie_fig}
\end{figure}

Figure \ref{pixie_fig} shows the PIXIE instrument concept.
It consists of a polarizing Fourier transform spectrometer (FTS)
with two input ports illuminated by co-pointed beams on the sky.
A set of polarizing wire grids
splits each beam into orthogonal liner polarizations
then mixes the beams.
A pair of movable mirrors
introduces an optical phase delay
before a second set of polarizing grids
recombines the beams
and routes them to a pair of
polarization-sensitive detectors
at each of the two output ports.
As the phase-delay mirrors sweep back and forth,
each of the 4 detectors samples the resulting 
interference fringe pattern
as a function of the optical phase delay.
Let $\vec{E} = E_x \hat{x} + E_y \hat{y}$ 
represent the electric field incident from the sky.
The power $P$ at the detectors
as a function of the phase delay $z$
may be written
\begin{eqnarray}
P_{Lx} &=& 1/2 ~\smallint \{ ~(E_{Ax}^2+E_{By}^2)+(E_{Ax}^2-E_{By}^2) \cos(z\omega /c) ~\}d\omega   \nonumber \\
P_{Ly} &=& 1/2 ~\smallint \{ ~(E_{Ay}^2+E_{Bx}^2)+(E_{Ay}^2-E_{Bx}^2) \cos(z\omega /c) ~\}d\omega   \nonumber \\
P_{Rx} &=& 1/2 ~\smallint \{ ~(E_{Ay}^2+E_{Bx}^2)+(E_{Bx}^2-E_{Ay}^2) \cos(z\omega /c) ~\}d\omega    \nonumber \\
P_{Ry} &=& 1/2 ~\smallint \{ ~(E_{Ax}^2+E_{By}^2)+(E_{By}^2-E_{Ax}^2) \cos(z\omega /c) ~\}d\omega~,
\label{full_p_eq}
\end{eqnarray}
where
$\hat{x}$ and $\hat{y}$ refer to orthogonal linear polarizations,
L and R refer to the detectors in the left and right concentrators,
A and B refer to the two input beams,
and $\omega$ is the angular frequency of incident radiation.
The optical phase delay $z$ is related to the physical mirror position $\Delta L$ as
\begin{equation}
z = 4 \cos(\theta) \cos(\delta/2) \Delta L ~,
\label{mirror_eq}
\end{equation}
where
$\theta$ is the angle of incident radiation
with respect to the mirror movement,
$\delta$ is the dispersion in the beam,
and the factor of 4 
reflects the symmetric folding of the optical path.
The factor of $1/2$ rather than $1/4$ for each of the four detectors
results from use of 2 input ports
rather than a single port.
When both input ports are open to the sky,
the power at each detector consists of a dc term 
proportional to the intensity 
$E_x^2 + E_y^2$
(Stokes $I$)
plus a term 
modulated by the phase delay $z$,
proportional to the linear polarization 
$E_x^2 - E_y^2$
(Stokes $Q$) 
in instrument-fixed coordinates.
Rotation of the instrument about the beam axis
rotates the instrument coordinate system
relative to the sky
to allow separation of Stokes $Q$ and $U$ parameters on the sky.
A full-aperture calibrator can be deployed
to block either of the two input ports,
or stowed so that both ports view the sky.
With one input port terminated by a blackbody calibrator,
the modulated term is then proportional to the
difference between the sky signal 
and the calibrator,
providing sensitivity to the sky signal in Stokes $I$, $Q$, and $U$
as well as a known reference signal for calibration.
Scattering filters on the folding flat
and secondary mirrors
limit the optical passband
to truncate the integral in Eq. \ref{full_p_eq}.
The 30~$\mu$m wire spacing in the polarizing grids
additionally limits the response to high-frequency signals.
Both effects are included in all calculations below.

\subsection{PIXIE Calibrator}

\begin{figure}[b]
\centerline{
\includegraphics[width=6.0in]{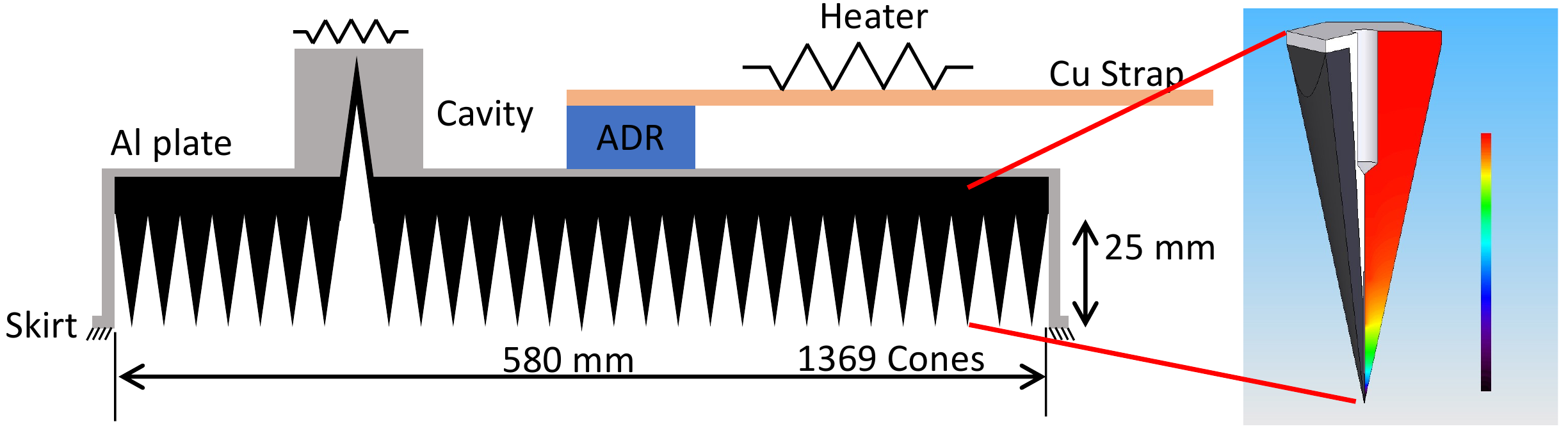}}
\caption{
Schematic of PIXIE blackbody calibrator with inset showing 
front-to-back thermal gradients in a single cone.
}
\label{xcal_fig}
\end{figure}

Figure \ref{xcal_fig} shows the calibrator design.
It will consist of 1369 absorbing cones
mounted on an aluminum backplate 580 mm in diameter.
Each cone is 25 mm tall and 16 mm wide at the base,
with an aluminum core coated by an absorbing layer.
The absorber consists of a combination of
graphite, stainless steel, and doped silicon particles
5 $\mu$m in diameter
suspended in a matrix of epoxy and silicon dioxide powder
\cite{steelcast_2008}.
The matrix is tuned so that its coefficient of thermal expansion
matches the aluminum core.
The aluminum core provides thermal conductivity
and mechanical support
while allowing easy attachment to the aluminum back plate.
Some 40 thermometers mounted in selected cones
monitor the temperature across the calibrator.

Thermal control is maintained using three elements.
A copper strap runs from the calibrator
to a heat sink maintained at 2.6 K
to provide the main cooling path for the calibrator.
A resistive heater allows for precise thermal control.
A paramagnetic salt pill within a controlled magnetic field
acts as a single-stage adiabatic demagnetization refrigerator
to provide a reversible source of heat.
The magnetic field can be reduced to cool the calibrator
below the 2.6~K heat sink,
or the field can be increased
to heat the calibrator well above the heat sink.
Since the heat is stored within the salt pill,
raising or lowering the calibrator temperature
in this fashion
does not inject an additional heat load to the 2.6~K heat sink.

Both the calibrator and the instrument optical path
(mirrors, polarizing grids, and surrounding walls)
will normally 
be maintained within a few mK of the CMB monopole temperature,
approximating a blackbody cavity.
A blackbody is fully characterized by a single parameter,
its temperature:
in the limit that the instrument and sky are fully isothermal,
no fringe pattern can be produced
regardless of the position of the mirrors
or the emissivity of any portion of the instrument.
Deviations from this ideal,
whether from the sky or the instrument,
will then generate non-zero interferograms
(Eq. \ref{full_p_eq}),
but by approximating an isothermal cavity,
PIXIE reduces the  required dynamic range
and suppresses instrumental effects
(systematic errors)
from all components following the FTS
\cite{nagler/etal:2015}.

A blackbody calibrator at temperatures near 2.725 K
provides an excellent match to the CMB,
but emits almost no radiation at frequencies above 600 GHz
where other astrophysical sources
(interstellar dust, 
the cosmic infrared background,
and far-infrared emission lines)
become important.
Simply raising the calibrator temperature to 20~K
would provide photons 
for the high-frequency calibration,
but would also increase the total power on the detectors
by a factor of over 1000.
Non-linearities in the detector performance
would then limit the reliability of the resulting calibration.
Instead, PIXIE replaces a single cone
with a matching conical cavity
which can be heated to higher temperatures.
The filling fraction of a single cavity (of order 0.1\%)
approximates the effective emissivity
of the dominant dust foreground,
while the choice of a cavity instead of a cone
minimizes perturbations from radiative heat transport
between the cavity and surrounding cones.
Normally the cavity is maintained at the same temperature
as the rest of the calibrator.
Section 2.2 discusses the high-frequency calibration
when the cavity is heated above the CMB temperature.

The PIXIE calibrator design is based on a similar calibrator 
flown by the ARCADE-2 mission
\cite{arcade_cal_2006},
which provides a more compact configuration
than the single inverted cone
flown by FIRAS
\cite{firas_cal_1994}.
The ARCADE-2 calibrator achieved
reflectance between -42 dB at the longest wavelength of 10 cm (3 GHz)
and -68 dB at shorter wavelengths,
averaging -57 dB across the full band 30--90 GHz
\cite{arcade_cal_2006}.
The PIXIE calibrator is expected to achieve reflectance
less than $-65$ dB from 15 GHz to 600 GHz,
rising to $-50$ dB at higher frequencies.
In the long-wavelength limit,
the calibrator may be approximated
as a smooth impedance transition
from free space to the absorber.
The ARCADE-2 calibrator used 295 cones each 88 mm tall,
requiring the transition to occur over less than one wavelength
at the lowest frequency of 3 GHz.
For PIXIE at 30 GHz, the transition takes place over 2.5 wavelengths.
The measured reflectance of the ARCADE-2 calibrator
at a comparable ratio of cone height to wavelength
is -68 dB (the noise floor of the measurement).
In the short-wavelength (geometric optics) limit,
the array of cones can be approximated as a series of surfaces
requiring multiple bounces before incident rays can reflect back.
Critical parameters
are the dielectric loading of the absorber,
the number of bounces,
and
the sharpness of the cone tips 
and joints between adjacent cones.
The ARCADE-2 short-wavelength reflectance of -56 dB
is consistent with
6 bounces 
with $\sim$12\% reflection at each bounce.
The PIXIE calibrator 
retains the same number of bounces but
tunes the absorber dielectric constant and loading
to reduce the surface reflectance at each bounce.
Replacing the 
alumina (Al$_2$O$_3$) filler in the ARCADE-2 calibrator
with
silicon dioxide (SiO$_2$) for PIXIE
reduces the index of refraction in the absorber mixture,
lowering the surface reflection by 40\%.
The limiting factor is expected to be 
single reflections from the cone tips.
The ARCADE-2 cones have a tip radius of 300 $\mu$m.
Engineering models of the PIXIE cones
demonstrate tip radius below 50 $\mu$m,
corresponding to calibrator reflectance -50 dB
at frequencies above 600 GHz
(wavelength $< 500~\mu$m).
Reflectance is further minimized
by staggering the heights of individual cones
so that reflections do not add coherently
and by
tipping the entire calibrator by $\sim$2\deg~
so that residual reflections
do not propagate directly back 
through the instrument to the detectors
($\S$3.3).

Temperature gradients within the calibrator
are expected to be small.
Steady-state thermal gradients require heat transport within the calibrator.
The copper strap attaches to the calibrator at a single point:
heat transport within the calibrator requires heat to flow
from the strap through the absorbing cones 
and then by radiation to the instrument optics.
We estimate the resulting gradients
by comparison to the ARCADE-2 calibrator.
The balloon-borne ARCADE-2 calibrator 
operated at altitude 37~km
and was surrounded by gaseous helium at pressure $\sim$3~Torr.
Temperature control at 2.725~K was maintained by heaters mounted
on the back of the calibrator,
while the absorbing cones faced metal structures
(conical horn antennas and mounting plates)
at temperature 1.5~K.
Heat flow from the heaters
through the absorbing cones
and helium gas
to the colder metal structures
drove back-to-front temperature gradients
within the absorbing cones.
In-flight data
showed heat flow of 1.5~W
resulting in 600 mK temperature gradients
concentrated near the cone tips
\cite{fixsen/etal:2011}.
The PIXIE calibrator,
in contrast,
operates in a vacuum
with the absorbing cones
facing a nearly isothermal cavity.
Radiative transport between the calibrator at 2.725~K
and a cavity 5~mK colder
produces a heat flow of 0.3~nW,
corresponding to maximum back-to-front temperature gradients
of 0.1~nK within the absorbing cones
(distortion $|y| \sim 10^{-12}$).
This is smaller than the thermometer read noise ($\S$3.1)
even when averaged over the course of a four-year mission.

In the limit that the calibrator is an open thermal circuit,
it cannot support thermal gradients.
Energy deposition from cosmic rays traversing the calibrator
represent a stochastic internal source of heat
to produce thermal gradients across the calibrator.
The flux of cosmic ray protons 
has a steep energy dependence, 
$dN/dE \approx 0.003~(E/10~{\rm GeV})^{-2.7}$
cm$^{-2}$~sr$^{-1}$~s$^{-1}$~GeV$^{-1}$
at energies above a few GeV,
yielding an anticipated rate of order 1 hit per second
for a single cone\footnote{
The cosmic ray flux at lower energies is suppressed
by the heliosphere.}.
A single cosmic ray hit will deposit $\sim$50~MeV (8 pJ) 
into the calibrator,
nearly independent of the proton energy.
Each cone has heat capacity $5 \times 10^{-3}$~J/K.
In the absence of thermal averaging a single proton hit would
instantaneously raise the temperature of a cone by 2 nK,
corresponding to distortion $|y| < 10^{-12}$
after accounting for signal dilution from other unaffected cones.
In practice, we must account 
for both the thermal time constant
and for multiple cosmic ray hits over the entire calibrator.
Averaged over the entire calibrator,
we expect $\sim$1000 cosmic ray hits per second,
equivalent to an internal heat dissipation of 5~nW.
The resulting flow of heat 
through the calibrator base to the copper strap heat sink
creates a radial temperature gradient of order 1 nK across the calibrator,
equivalent to a distortion $|y| < 10^{-10}$.

Temperature gradients 
during the high-frequency calibration
using the single hot cavity
are more difficult to assess.
A single calibrator cavity with diameter 16 mm
and temperature 20~K
will radiate $\sim 0.5~\mu$W
to the 2.725~K instrument,
with negligible radiation
from the instrument back to the cavity.
If gradients within the calibrator cavity
scale comparably to gradients in the absorbing cones,
the resulting radiative heat flow
would drive back-to-front gradients of order 0.2~mK
within the absorbing material of the cavity.
Lateral gradients along the cavity walls
and from the cavity to nearby cones
are likely to be larger than this.
A 2 mK lateral gradient within the cavity
corresponds to fractional error
of order 0.01\%
in the absolute calibration
above 600 GHz.
As with ARCADE,
lateral gradients in the cones surrounding the cavity
can be identified and removed
using a principal component analysis
of transients excited
as the cavity is commanded to different temperatures.
A principal component analysis 
reduced the
measured 600~mK front-to-back thermal gradients in the ARCADE-2 calibrator
to a 5~mK uncertainty in the 
calibrator radiometric temperature
\cite{fixsen/etal:2011}.
A comparable analysis for PIXIE 2~mK gradients
would predict uncertainties of order 20~$\mu$K
in the high-frequency calibration.
Additional analysis of temperature gradients
deliberately induced when the calibrator is stowed
can further constrain 
calibrator time constants and thermal profile.

\subsection{Calibration Operations}

PIXIE will observe from the second Sun-Earth Lagrange point, L2.
The phase-delay mirrors complete a full cycle every 3 seconds.
The co-pointed beams are maintained perpendicular
to the Sun-Earth line.
The entire observatory rotates about the beam axis
every 48 seconds
while simultaneously scanning the beams through a great circle
perpendicular to the Sun-Earth line
every 384 rotations
(approximately 5 hours).
The annual motion of the Sun-Earth line 
precesses the great circle on the ecliptic plane once per year
so that the beams map the full sky 
every 6 months.
The fixed relations of 
16 mirror cycles per beam rotation
and 
384 beam rotations per great-circle scan
simplify the mapping algorithm  
but are not required.


Calibration proceeds as follows.
At the start of each great-circle scan,
the calibrator is set to a new temperature
and moved to a new position.
Temperature setpoints span the range [2.720, 2.730]~K
to bracket the CMB monopole at 2.725~K.
The calibrator position sequentially deploys
through 4 consecutive scans
to block the A beam,
leave both beams open to the sky,
block the B beam,
and leave both beams open to the sky again.
This pattern gives equal integration time
to measurements of CMB polarization
(both beams open)
and spectral distortions 
(calibrator over one beam)
but can be modified at any time during the mission.
We define the start of a scan as either the north or south ecliptic pole,
so that changes in temperature and position
occur only over regions of the sky observed at every scan.
Temperatures for other elements of the instrument
can also be changed at the start of each scan
($\S$3.3).

With the calibrator deployed to block one beam,
the interferograms depend on the difference between
the sky signal and the calibrator.
We remove dependence on the sky signal
by comparing data from two neighboring scans:
one with the calibrator blocking the A beam
and a second with the calibrator blocking the B beam.
Since input signals change sign
when observed in the A beam vs the B beam
(Eq. \ref{full_p_eq}),
the sky signal in each map pixel
cancels when summing the scan data for that pixel,
leaving only the difference in the calibrator signal\footnote{
The maximum great circle motion 
of 0.4\deg~between the two calibration scans
is small compared to the 2.2\deg~tophat beam
or the 0.9\deg~pixel diameter.
Appropriate interpolation techniques further reduce 
the impact of scan precession.
The differential optics cancel effects of non-ideal beam shape
(ellipticity, etc) 
prior to detection.
Coupling errors between beam and sky are further reduced
by Fourier transform with respect to rotation angle
\cite{kogut/etal:2019}.
}.
The differential calibrator signal
may then be averaged over the scan circle
to provide an absolute blackbody reference
for signal calibration. 
Transients following commanded temperature changes
and
pixels with large sky-signal gradients
(e.g. near the Galactic plane)
may be excluded from the calibration
with only modest loss of effective observing time.


Calibrator excursions of $\pm$5 mK about the mean CMB temperature
provide calibration for frequencies below 600 GHz.
A small fraction of the observing time (5--10\%)
is reserved for operations
with the entire calibrator at significantly 
higher or lower temperatures,
or with the single inverse cavity heated to higher temperature.
These data provide calibration for data above 600 GHz,
and will be supplemented by observations of Jupiter, Mars, and Saturn,
each of which is observed for 13 consecutive scans twice per year.
The total power absorbed by each detector
is dominated by the CMB monopole and the calibrator mean temperature;
the fringe pattern from 5 mK calibrator excursions
changes the detector loading by only 0.7\%
to minimize effects of detector non-linearities
for the low-frequency calibration.
When the single cavity is heated to 20~K,
the power absorbed by each detector nearly triples.
To compensate for this,
the temperature of the rest of the calibrator
can be lowered 
so that the total power on the detector remains nearly constant.
Comparison of scans at identical detector loading
but different cavity temperatures
then determines the filling fraction of the single cavity.
Selected scans with the entire calibrator
maintained at temperatures ranging from 2.5 to 3~K
vary the detector loading by 50\%
to provide channel-by-channel
characterization of detector non-linearity
so that the calibrated sky spectra
are always interpolated
and not extrapolated.

\section{Calibration Accuracy}

\begin{figure}[t]
\centerline{
\includegraphics[width=5.0in]{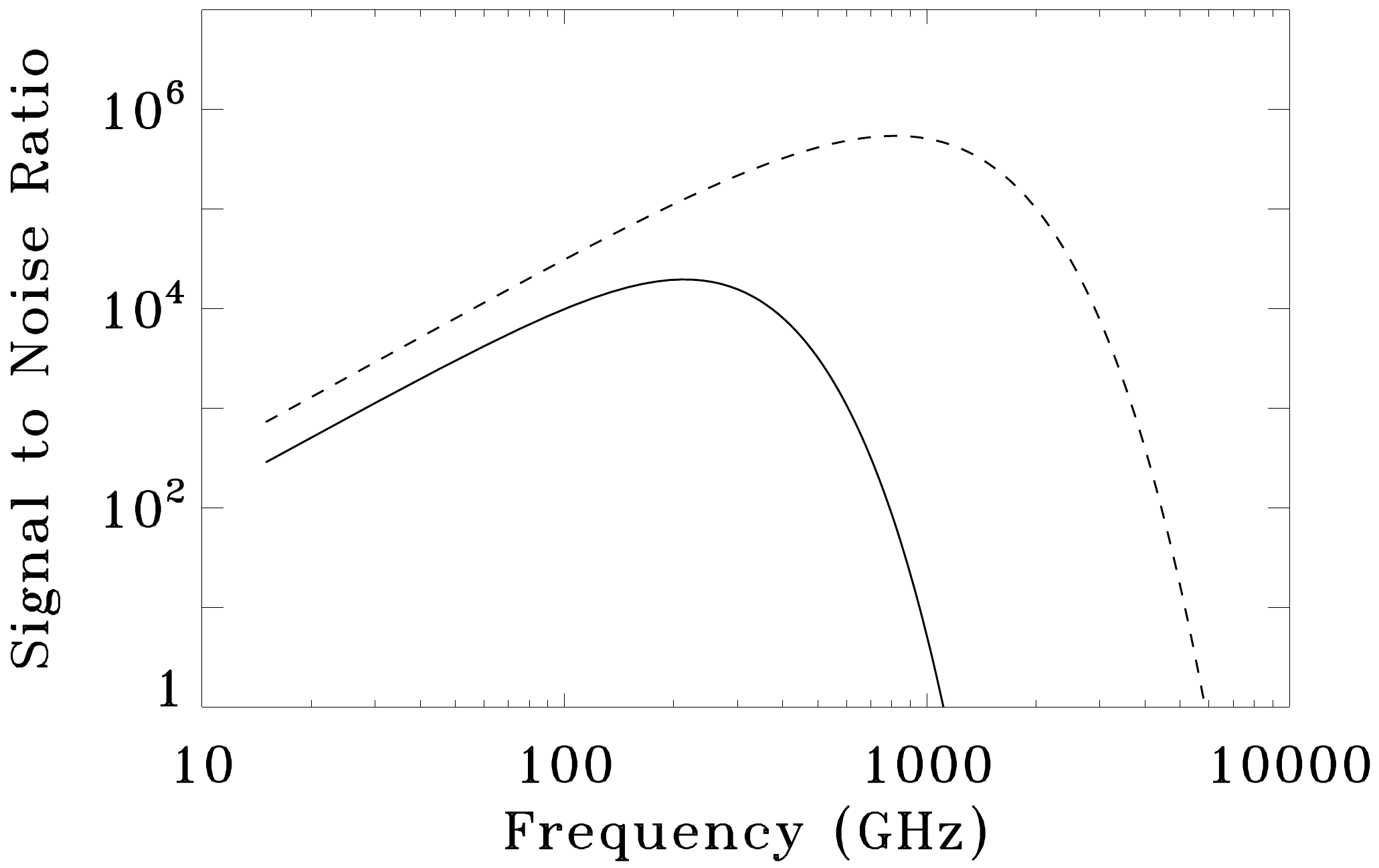}}
\caption{
Signal to noise ratio for a single differential calibration scan.
The solid line refers to the low-frequency calibration
with the entire calibrator maintained within 5 mK
of the CMB temperature.
The dashed line shows the high-frequency calibration
with a single cavity at 20~K.
The PIXIE calibration will not be limited by signal to noise ratio.
}
\label{cal_snr_fig}
\end{figure}

The accuracy to which sky signals can be converted
from digitized telemetry units to physically meaningful units
depends on the signal to noise ratio of the calibration,
the accuracy of the temperature scale,
and the extent to which non-ideal instrument parameters
(reflections, temperature gradients) can be removed.
PIXIE will be background-limited,
with noise dominated by photon arrival statistics
from the CMB and blackbody calibrator.
Calibrator temperature excursions of $\pm$5 mK
provide a signal-to-noise ratio 
above $10^3$
within individual frequency channels below 600 GHz
for a single comparison of two neighboring calibrator scans
(Figure \ref{cal_snr_fig}).
Operations with a single hot cavity provide
signal to noise ratio up to $10^6$
at higher frequencies.
Raw sensitivity will not limit the PIXIE calibration.
The following sections discuss the impact 
of various instrumental effects on the PIXIE calibration.

\subsection{Thermometry}

The PIXIE calibration rests upon comparison of blackbody loads
at different temperatures.
The FTS and beam-forming optics,
including all walls and baffles,
will be maintained at temperatures near the CMB monopole,
allowing the temperature readout to be optimized
for a relatively narrow range.
Thick-film ruthenium oxide resistors
with resistance $\sim$50 k$\Omega$ at 2.7~K
and 1~pW power dissipation
can be read out to precision
$\sim 100~\mu$K
over a one-second integration.
A principal component analysis of the ARCADE-2 calibrator
showed that 7 well-placed thermometers
would be sufficient to characterize
both the mean temperature 
and thermal gradients within the calibrator\cite{fixsen/etal:2011}.
The PIXIE calibrator will contain 40 thermometers
mounted at different heights
in individual cones.
The pW power dissipation from each thermometer is small 
compared to the nW front-to-back heat flow
within the calibrator
and will not contribute significantly to thermal gradients.
Thermometry noise
limits knowledge of the calibrator mean temperature
to precision 16~$\mu$K for a one-second integration,
or 100 nK for a single differential calibration scan.
A four-year mission provides over 1600 calibration scans,
reducing the thermometry noise to the 2 nK level
over the entire mission,
corresponding to calibration uncertainty
of order 
$10^{-9}$
within individual frequency channels.
The resulting spectral distortion limits
$|y| < 2 \times 10^{-11}$
and
$|\mu| < 3 \times 10^{-10}$
are negligible compared to the instrument noise.

The one-second thermometer sampling is rapid
compared to the 20 second thermal time constant
of the calibrator.
Thermal excursions on shorter time scales
({\it e.g.} from energy deposited
by cosmic ray hits within the calibrator)
are suppressed by 
the thermal mass
and filling fraction of the absorbing cones
($\S$2.1).
The expected cosmic ray flux
yields stochastic fluctuations 0.2~nW
over the course of a 1-second thermometer integration,
producing negligible distortions in the calibrator-sky difference spectra.

Careful distinction must be made between
thermometer precision vs accuracy.
Although the thermometer readout noise 
integrates down with additional calibration scans
(precision),
the conversion between thermometer resistance
and absolute temperature (accuracy)
does not.
A well-calibrated cryogenic thermometer
will typically reproduce the actual temperature within 100~$\mu$K,
leading to comparable uncertainty in the
physical temperature of the calibrator.
Different calibration scans 
at different commanded calibrator temperatures
bracket the CMB monopole temperature.
Over this limited range,
the absolute thermometry error
may be treated as a constant temperature offset,
canceling when comparing calibrator data
at different commanded temperatures.
By interpolating sky spectra taken
at different calibrator temperatures,
PIXIE can match the (unknown) monopole temperature
to 2 nK precision;
however,
the inferred monopole temperature
will differ from the true temperature
by up to 100 $\mu$K.
Since the sky signal does not enter the calibration process,
this systematic error does not affect the relative calibration.

The absolute thermometry does not depend solely on pre-flight calibration,
but can be deduced from in-flight data as well.
Calibration data from successive scans
with calibrator temperature difference $\Delta T$
generate differential spectra
\begin{equation}
I(\nu,T) = \frac{\partial B_\nu}{\partial T} \Delta T
\label{dbdt_eq}
\end{equation}
where
\begin{equation}
B_\nu(T) = \frac{2 h \nu^3}{c^2}
	   \frac{1}{{\rm e}^x - 1}
\label{planck_eq}
\end{equation}
is the blackbody Planck law,
$\nu$ is the observing frequency,
$k$ is Boltzmann's constant,
$h$ is the Planck constant,
and
\begin{equation}
x = \frac{h \nu}{k T}
\label{x_eq}
\end{equation}
is the dimensionless frequency,
defined using the thermodynamic temperature $T$.
We may fit the peak frequency of the observed calibrator spectra\footnote{
Since PIXIE does not directly measure 
a monopole spectrum,
but only the difference spectrum 
between the two input ports,
this method
uses the derivative of the Planck law
rather than the monopole spectrum itself.}
and use the Wien displacement law
to determine the mean calibrator temperature
to 10~$\mu$K accuracy
independent of the sky spectra,
corresponding to calibration accuracy $2 \times 10^{-5}$.
Alternatively,
we may use the sky-only data
to determine the color temperature of the
CMB dipole induced by PIXIE's motion about the sun,
which fixes the temperature scale to similar accuracy.
Both methods can assess thermometry drifts in time.

The 10~$\mu$K systematic error in the assigned temperature
will propagate to CMB spectral distortions,
which use the CMB monopole temperature
to define the distortion spectral shape.
This effect is small.
The optically-thin $y$ distortion is defined as
\begin{equation}
I(\nu) =  \frac{2 h \nu^3}{c^2}
\frac{1}{ {\rm e}^x - 1} \left(
	1 + y \frac{x {\rm e}^x}{ {\rm e}^x - 1}
	\left[ \frac{x}{ {\rm tanh}(x/2) } - 4 \right] \right) ~,
\label{y_eq}
\end{equation}
while the optically thick $\mu$ distortion is
\begin{equation}
I(\nu) =  \frac{2 h \nu^3}{c^2}
\frac{1}{ {\rm e}^{x + \mu} - 1 }
\label{mu_eq}
\end{equation}
A 10 $\mu$K error in the assigned value for $T_0$
creates a negligible fractional error 
$\Delta I/I < 2 \times 10^{-5}$
in the recovered $y$ or $\mu$ distortions.

\subsection{Temperature Gradients}
Temperature gradients within the calibrator
produce a superposition of emission
from elements at different physical temperatures.
The predicted 1~nK radial gradient
or
0.1~nK front-to-back gradient
produce negligible errors
in the absolute calibration:
even if the entire calibrator
were 1~nK warmer or colder than commanded,
the fractional error in the absolute calibration
is below $10^{-8}$.
However,
if not accounted for,
the resulting distortion 
$y \sim 10^{-10}$
from the calibrator itself
would induce a false detection
of opposite sign in the calibrator--sky comparison.
Radial gradients can be identified and subtracted
using a principal component analysis
during commanded calibrator temperature changes\cite{fixsen/etal:2011}.
The sign of the front-to-back temperature gradient 
depends on the sign of the calibrator--cavity temperature difference,
with the cone tips colder than the base
when the calibrator is warmer than the cavity
and warmer than the base
when the calibrator is colder than the cavity.
Since the calibrator spends equal amounts of time 
warmer as colder
over the course of the full mission,
thermal gradients cancel 
in the determination of the mean spectral distortion signal
and only contribute to the scan-to-scan variance.
The predicted effect is negligible compared to the thermometer readout noise.

Gradients in the single hot cavity affect the high-frequency calibration
by altering the power incident on the detector.
Since these data are used only as a reference for
sky signals above the CMB Wien cutoff,
they do not directly source a false $y$ distortion.
A 2 mK cavity gradient
creates a fractional error $7 \times 10^{-5}$
in the calibration at frequencies above 600 GHz.

\subsection{Reflections}

Internal reflections
within the calibrator--instrument system
create calibration errors
by replacing emission from the calibrator
with emission from elsewhere within the instrument.
Rays that reflect from the calibrator
and terminate elsewhere within the instrument
create an error signal proportional to
the calibrator power reflection coefficient
and the temperature difference
between the calibrator
and the instrument surface
where the ray terminates.
Most of the instrument
is maintained within 5 mK of the calibrator temperature
to minimize this effect.
A -65 dB calibrator reflection
results in a 1.5~nK error signal,
corresponding to a fractional error $3 \times 10^{-9}$
in the absolute calibration.

As with temperature gradients within the calibrator,
reflections to components at different physical temperatures
lead to a small $y$ distortion in the calibration signal.
Each of the $\sim$20 surfaces 
in the cryogenic portion of the instrument
is individually temperature controlled.
The commanded temperatures are periodically varied
to allow identification and subtraction
of the residual reflected signals,
which also maps beam spillover
within the instrument
\cite{nagler/etal:2015}.
The calibration procedure includes
$\sim$20 additional free parameters
for the coupling to each internal surface.
This number is small compared to the 40 million mirror strokes
during the course of the mission,
allowing the residual $y$ distortion
to be removed to the mission noise limit
$|y| < 10^{-11}$.

The detector assembly is maintained at 0.1~K
and is the only component significantly colder 
than the calibrator.
Approximately 0.3\% of the 
signal reflected by the calibrator
(-90~dB of the full signal)
propagates back through the FTS
to reach the detector.
The corresponding 3~nK error signal
may be removed to first order
using pre-flight measurements of the 
calibrator reflection,
leaving residuals
$|y| < 10^{-11}$ in the distortion spectra.

The reflection error signal is larger
for the high-frequency calibration
using the single hot calibrator cavity.
Since the 2.725~K instrument
has negligible emission above 600 GHz,
the fractional error in the high-frequency calibration
primarily depends on the power reflection coefficient 
of the single hot cavity.
A -40~dB reflection from the cavity
produces fractional errors of $2 \times 10^{-5}$
in the high-frequency calibration.

\subsection{Leakage}

Leakage of external signals into the calibrator/instrument cavity
presents an additional perturbation.
The CMB monopole dominates the sky at frequencies below 600 GHz.
Since the calibrator is maintained within a few mK of the CMB,
sky leakage presents a negligible problem.
The calibrator and instrument optics
are surrounded by a nested set of reflective shields,
preventing direct view from the instrument aperture
to the Sun or warm portions of the spacecraft.
Potential leakage is dominated by the view
to the innermost shield,
which operates at temperatures near 16~K
and subtends 3\% of the $2\pi$~sr above the aperture.
To minimize leakage,
the calibrator extends 2~mm beyond the edge of the aperture
with a flexible skirt of aluminized mylar
to provide an electromagnetic seal
between the calibrator and the aperture.
The barrel has a flared top to apodize the aperture,
placing the actual interface between the calibrator and the aperture
beyond direct view of the detectors.
The estimated leakage of 70~dB
produces an error below $10^{-8}$
in the absolute calibration.

\section{Frequency Calibration}

Fourier transform spectroscopy allows a single detector
to produce data at a large number of
well-characterized frequency channels.
Let 
$S_\nu$ represent the frequency-dependent sky signal
and
$S_k$ represent the amplitude of the sampled fringe pattern.
The two are related by a Fourier transform, 
\begin{eqnarray}
S_k   &=& \int S_\nu \exp\left( \frac{2\pi i z_k \nu}{c} \right) ~{\rm d}\nu~,	
\qquad
S_\nu 
=
\sum_{k=0}^{N_s-1}{ W_k \, S_k \exp\left( \frac{2\pi i \nu k Z}{c N_s} \right)}~,
\label{syserr_fourier_eq}
\end{eqnarray}
where
$\nu$ denotes frequency,
$z_k$ is the optical phase delay for fringe sample $k$,
$W_k$ is the apodization weight,
and
$k$ labels the individual fringe samples.
As the mirror moves,
we obtain $N_s$ detector samples
over an optical path length $\pm Z$.
The Fourier transform of the sampled fringe pattern
returns the sky signal at sampled center frequencies
$n ~ \times c/(2 Z)$
where
$n = 1, 2, ..., N_s/2$.
The maximum phase delay (optical stroke)
thus determines the width of the frequency bins
in the synthesized spectra,
while the number of detector samples within each optical stroke
determines the number of frequency bins
and hence the highest sampled frequency.

To facilitate subtraction of line emission,
the maximum optical phase delay $Z$ may be chosen
to be an integer multiple of the wavelength of the
$J=1-0$ CO line,
$Z = M (\lambda_{CO} / 2)$,
in which case
every $M^{th}$ synthesized channel
is centered on a CO line.
The PIXIE selection $Z = \pm10.4$~mm ($M=8$)
and
$N_s=1024$
yields
512 synthesized
channels from 
14.4~GHz to 7.3~THz\footnote{
To reduce optical loading and photon noise from zodiacal light,
scattering filters limit the optical passband
so that channels above a few THz contain no sky signal.}.
The corresponding mirror physical motion is $\Delta L = \pm 2.7$~mm
(including corrections for dispersion
and off-axis operation, Eq. \ref{mirror_eq}).
By appropriate choice of weights,
wider channel widths can be synthesized
to facilitate subtraction of other lines.
Radial velocities of individual emitting regions
will shift the observed line emission 
with respect to the rest frequency,
but this effect is small:
the $\sim$200~km~s$^{-1}$ Galactic rotation
shifts the observed frequencies
by less than the synthesized channel width
over the entire PIXIE band.

The weights $W_k$ determine the channel-to-channel covariance,
allowing {\it a priori} specification 
of the passband for individual frequency channels.
The response to a delta function in the frequency domain
is simply the Fourier transform of the window function
(the weights $W_k$).
The simplest window choice
is a boxcar:
$W=1$ for mirror position $|z| < Z$
and $W=0$ otherwise.
The corresponding Fourier transform is the sinc function
$\sin(x) / x$,
where
$x = (\nu - \nu_0) (Z / c)$
and $Z$
is the maximum optical path change.
The channel width
is thus inversely proportional to the maximum travel of the mirror.

\begin{figure}[b]
\centerline{
\includegraphics[width=5.0in]{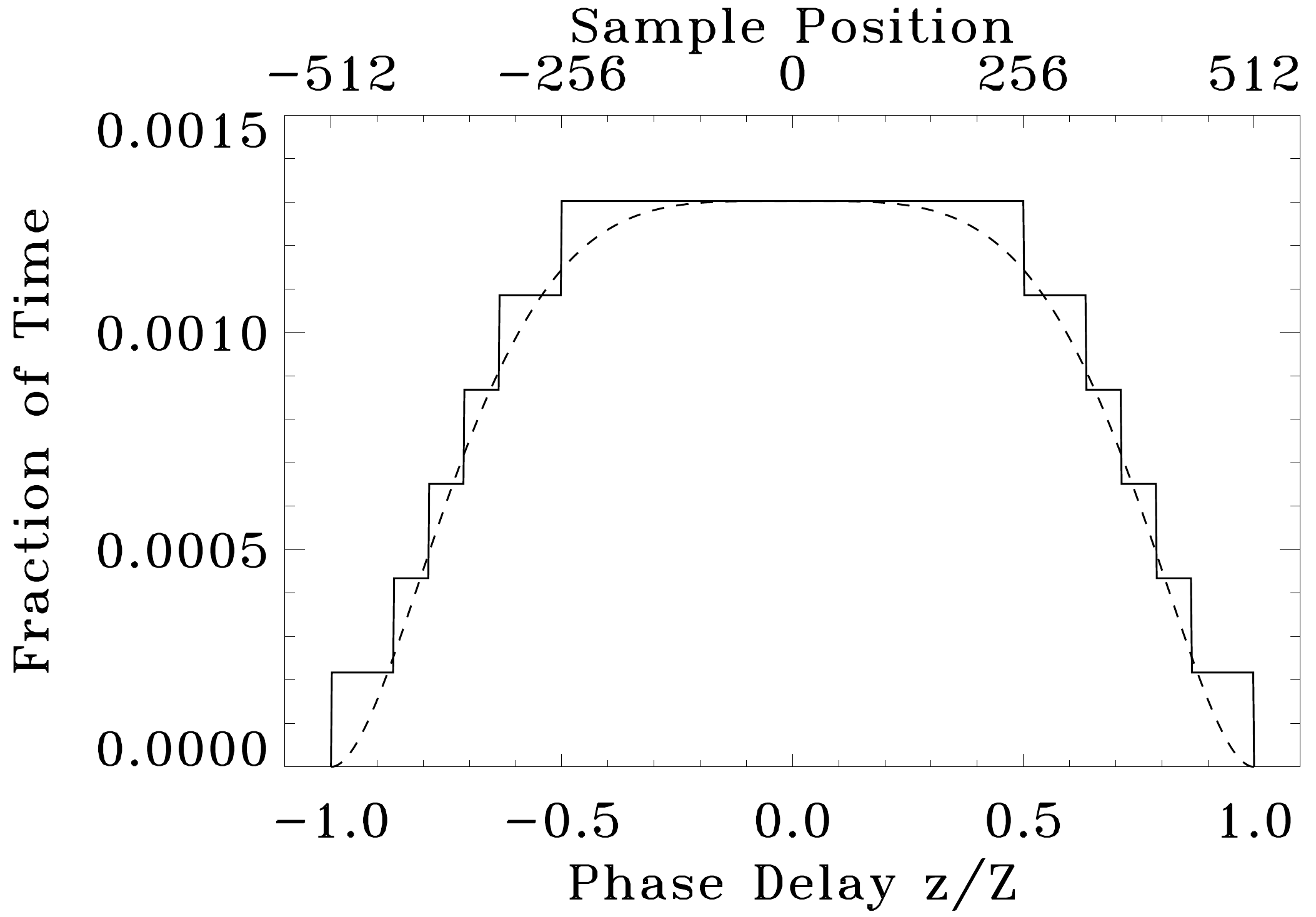}}
\caption{
The mirror stroke can be varied to apodize the sampled fringe pattern.
The dashed line shows the ideal apodization
(Eq. \ref{apodization_eq})
for a fringe pattern
sampled at 1024 mirror positions.
The apodization
achieved by varying the mirror start and stop position
in 6 fixed steps
(solid line)
accumulates samples to approximate the ideal apodization
without deweighting data
at large phase delay.
}
\label{apodization_fig}
\end{figure}

The boxcar window produces the highest frequency resolution,
but has several disadvantages for
continuum sources. 
It produces significant correlations (ringing)
over wide frequency ranges,
potentially aliasing signals
from outside the desired passband.
In addition,
the uniform weighting with respect to mirror position
gives too much weight to the far ends
and
too little weight near the white-light null
where most of the continuum information lies.
The solution is to apply an apodization function.
A wide variety of apodizations are in common use,
including trapezoid,
Hamming, Hanning, Gaussian, and cosine windows.
With suitable choice of weights $W_k$
any of these (or others)
can be employed even after the data are archived.
Here we folllow FIRAS
to use apodization
\begin{equation}
W(x) = (1-x^4)^2
\label{apodization_eq}
\end{equation}
as a smooth function that goes to zero at the ends,
where
$x = z/Z$
is the relative mirror displacement.
Although this apodization could be realized
using a uniform mirror stroke over the full range 
$x = [-1,1]$
and discarding (or deweighting) samples at larger $|x|$,
the integration time wasted on discarded samples
reduces the observing efficiency to 71\% of the ideal limit.
A straightforward implementation of this apodization
that retains high efficiency
would simply vary the mirror stroke,
with more strokes of short length
and relative few strokes over the full length,
approximating the desired apodization
without the need to discard (deweight) data 
near the maximum displacement.
To minimize dead time at turn-around,
we instead implement a fixed stroke length
$| \delta z | = 1.5 Z$
(75\% of the full end-to-end stroke length $2Z$)
and vary the start and stop positions
to approximate the desired apodization.
A fixed set of 6 start/stop positions
symmetric about zero phase delay
(Figure \ref{apodization_fig})
achieves 95\% observing efficiency
compared to the ideal apodization
(Eq. \ref{apodization_eq}).

PIXIE's frequency spectra are the Fourier transform
of the sampled fringe patterns.
Systematic errors in the phase delay (mirror position) $z$
create corresponding errors in the synthesized frequency channels.
Multiplicative errors in the mirror position
(e.g. from un-modeled thermal contraction)
create a matching scaling of the assigned frequencies.
Un-modeled optical errors
in the beam dispersion $\delta$
and illumination angle $\theta$
(Eg. \ref{full_p_eq})
will also affect the assigned frequencies.

PIXIE will use flight data to calibrate the frequency scale.
A simple offset $\Delta z$ shifts the white-light peak
at zero phase delay.
With over $10^8$ measurements of the white-light peak
over a 4-year mission,
a fixed offset is easily fitted and removed
to the noise limit of the full mission.

\begin{figure}[b]
\centerline{
\includegraphics[height=4.0in]{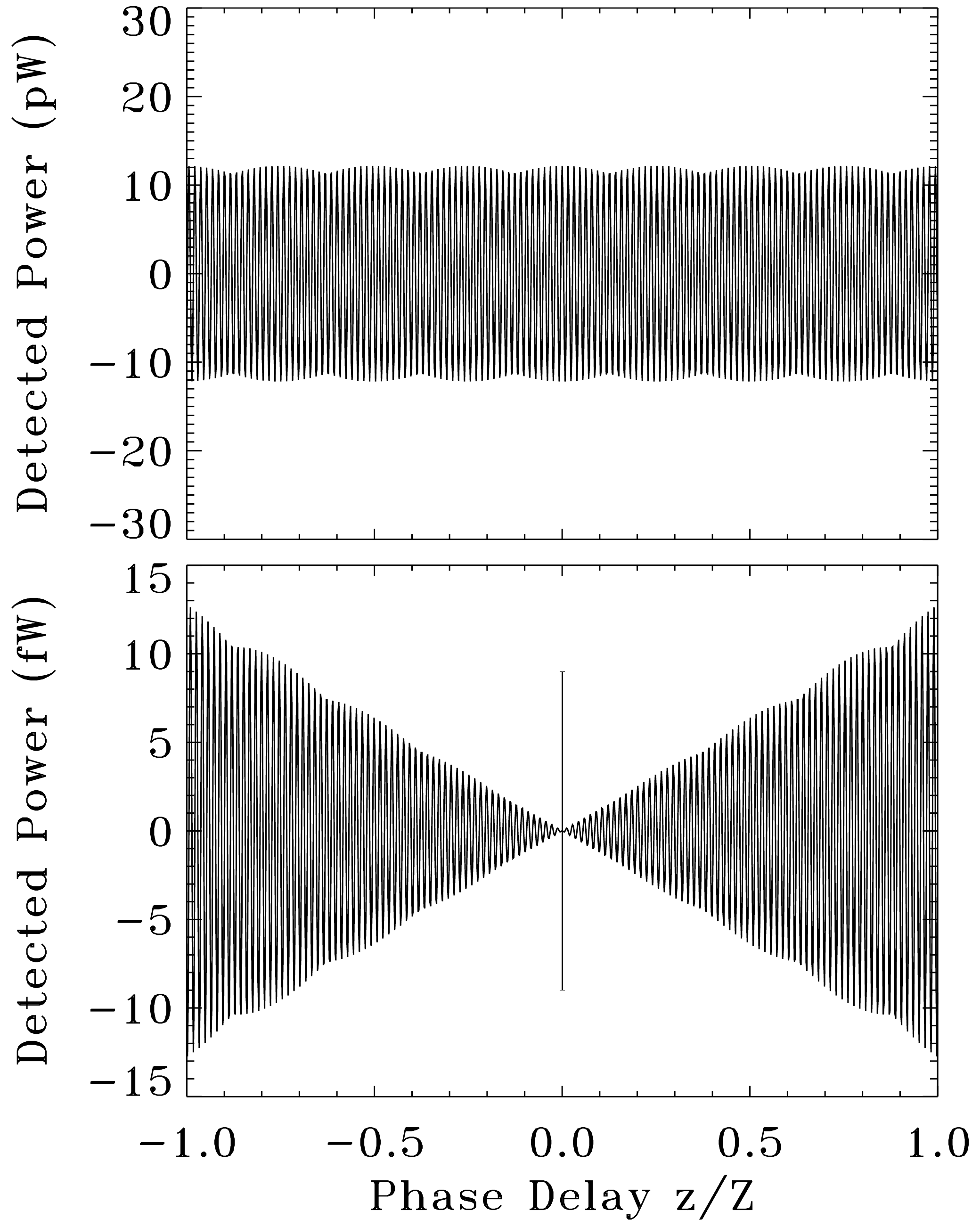}}
\caption{
Unresolved line emission produces a cosine fringe pattern
whose spatial frequency depends on the line optical frequency.
The top panels show the simulated fringe pattern 
from the median C{\sc ii} brightness
along the Galactic plane.
The bottom panel shows the signal change
produced by a 5 MHz shift in frequency
(note the units change from pW to fW).
The error bar at zero phase delay
shows the instrument noise for a single detector sample.
}
\label{cii_line_fig}
\end{figure}

Several methods determine the frequency scaling using in-flight data.
As in $\S$3.1,
we may use the Wien displacement law
to determine the frequency scale from the peak 
of the observed calibrator spectra.
The calibrator absolute temperature $T$
is known to 100~$\mu$K accuracy from  pre-flight calibration,
fixing the peak frequency of the observed spectra
to accuracy 6~MHz
(0.04\% of the synthesized channel width).
A similar fit to the peak frequency of the CMB dipole
and higher-order anisotropy
fixes the frequency scale to comparable accuracy
using sky data instead of calibration data
\cite{fixsen/etal:2009}.

Emission from bright unresolved Galactic lines
are Fourier transformed by the instrument 
to produce a simple cosine pattern in the observed fringe patterns.
In place of a simple Fourier transform
(Eq. \ref{syserr_fourier_eq}),
we may directly fit the fringe patterns to a cosine term,
thereby
determining the line frequency to higher precision
than the 14.4~GHz channel width of the synthesized spectra.
Figure \ref{cii_line_fig} shows the fringe pattern
for the bright C{\sc ii} line 
at rest frequency 1900.539 GHz.
The median C{\sc ii} intensity on the Galactic plane
produces fringes with peak amplitude 10~pW.
A frequency shift of 5 MHz
produces a differential signal
with amplitude 1000 times smaller,
detectable at 775 standard deviations
within a single mirror sweep.
Averaged over the full mission,
the C{\sc ii} line frequency
can be measured to 5 kHz accuracy
within each pixel along the Galactic plane.
The Galactic rotation curve is well mapped\cite{herschel_2013}:
a comparison of the fitted frequencies 
to the Galactic rotation curve
fixes the PIXIE frequency scale to accuracy 0.2 kHz.
A similar fit to line emission from the CO chain
$J=1-0$ through $J=7-6$
determines the frequency scale 
to comparable accuracy
at lower frequencies
overlapping the CMB
\cite{fixsen/etal:1996}.
A 200 Hz scale error at frequencies below 600 GHz
produces negligible error in the recovered sky spectra,
with both 
$\delta \mu / \mu $
and
$\delta y / y$
less than $10^{-8}$.

\section{Foregrounds and Dynamic Range}

The differential optical system
nulls the dominant monopole signal
but does not fully suppress foreground emission
whose spectra differ significantly
from the calibrator.
This does not affect the calibration
(which only requires the sky spectra to be constant in time),
but does place constraints on the instrument dynamic range.
The largest amplitude in the sampled fringe pattern
occurs at zero phase delay,
and is given by the integral of the difference spectra
between the two ports
(Eq. \ref{full_p_eq}).
With the calibrator deployed to block one beam,
the maximum fringe amplitude on a single detector
\begin{equation}
P_{\rm max} = 1/2 ~\int  
	\left[ {I^{\rm cal}(\nu) - I^{\rm sky}(\nu)} \right]
	 {A \Omega ~\alpha(\nu) ~\xi(\nu)} 
		                     d\nu
\label{pmax_eq}
\end{equation}
depends on the sky--calibrator differential spectrum,
where
$A \Omega$ is the etendue,
$\alpha(\nu)$ is the detector absorption coefficient,
and
$\xi(\nu)$ is the optical transmission efficiency
from the sky to the detector.

\begin{figure}[b]
\centerline{
\includegraphics[width=6.0in]{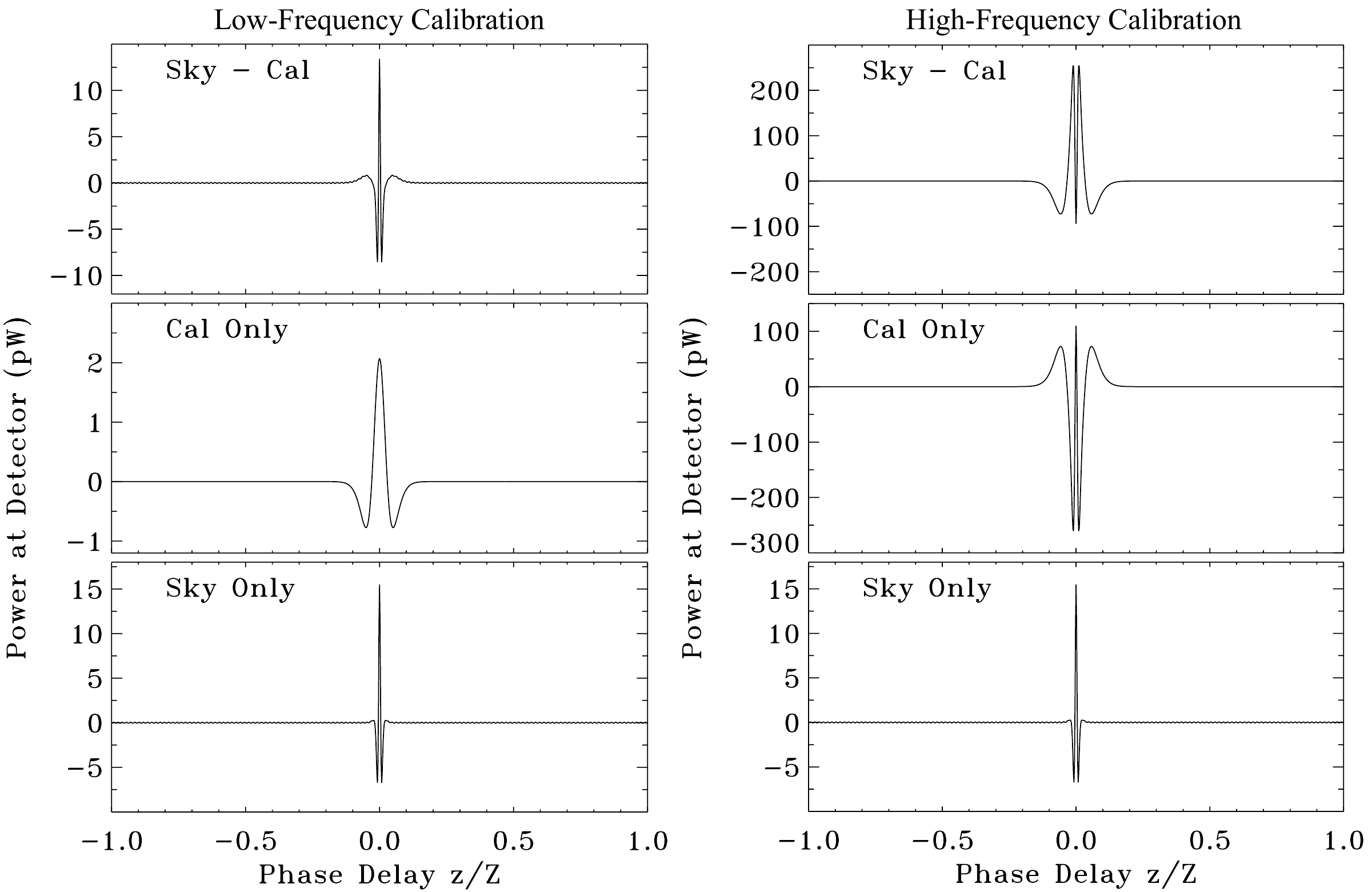}}
\caption{
Simulated fringe patterns for different sky / calibrator combinations
as the instrument views the median sky.
The top panels show the fringe pattern observed
when the calibrator covers one port while the other port
views the sky.
Center panels show the calibrator-only fringe pattern
produced by summing the signals over two adjacent calibration passes
($\S$2.2).
Bottom panels show the sky-only fringe pattern
from subtracting adjacent passes.
The left panels show the low-frequency calibration
with the entire calibrator maintained 5 mK above the CMB monopole.
The right panels show the high-frequency calibration
when the single cavity is heated to 20 K.
The detector noise in each plot is less than the line thickness.
}
\label{ifg_examples}
\end{figure}

Figure \ref{ifg_examples} shows a typical set of fringe patterns
during calibration.
We simulate the fringe pattern 
including continuum emission from the CMB,
Galactic synchrotron, free-free, thermal dust,
and the extragalactic contribution from the cosmic infrared background,
to which we add 
line emission from the bright 
C{\sc ii} $^2P_{3/2} \rightarrow ~ ^2P_{1/2}$
fine-structure line 
at 157.74~$\mu$m rest wavelength. 
We model synchrotron emission
as a power-law 
$I(\nu) \propto \nu^{\beta_s}$
with spectral index $\beta_s = -1.05$,
and model free-free emission
as a similar power law
with spectral index $\beta_{ff} = -0.16$.
We model thermal dust using a modified blackbody spectrum
$I_d \propto B_\nu(T_d) \nu^{\beta_d}$
with dust temperature 
$T_d = 19.6$~K
and spectral index
$\beta_d = 1.6$.
The synchrotron, free-free, and dust amplitudes
are normalized to the median brightness 
observed by Planck\cite{planck_fg_2018}.
We model the extragalactic infrared background
as a modified blackbody
$I_{\rm CIB} \propto B_\nu(T_{\rm CIB}) \nu^{\beta_{\rm CIB}}$
with
$T_{\rm CIB} = 18.5$~K,
$\beta_{\rm CIB} = 1.3$
and
amplitude normalization from FIRAS\cite{fixsen/etal:1998}.
The C{\sc ii} normalization
is set to 50\% of the median dust foreground
evaluated at the C{\sc ii} rest frequency.

When viewing the median sky
with the calibrator maintained 5~mK above the CMB monopole,
the sampled fringe pattern
peaks at amplitude $\sim$15~pW.
Summing signals from adjacent scans
with the calibrator at the CMB temperature vs 5~mK warmer
cancels the sky signal ($\S$2.2),
leaving just fringe pattern
from the 5~mK calibrator difference.
Similarly, subtracting signals from adjacent scans
leaves just the sky signal 
(minus the calibrator monopole and 5~mK difference signal).
The sharper peak in the sky-only signal
compared to the calibrator-only signal
results from diffuse dust emission,
while the smaller high-frequency ripples in the sky signal
result from line emission.
To calibrate emission at frequencies above the CMB Wien cutoff,
the single cavity in the calibrator is heated
while the main calibrator body is cooled.
The right panels in Figure \ref{ifg_examples} 
show the resulting high-frequency calibration.
Note that the summed calibrator-only signal
now contains sharply-peaked features
typical of higher-frequency continuum sources.

\begin{table}[t]
{
\small
\caption{Simulated Total Power and Fringe Amplitude}
\label{fringe_table}
\begin{center}
\begin{tabular}{c c c c }
\hline 
Calibrator	&	Cavity	&	Total	&	Fringe		\\
Temperature	&	Temperature &	Power	&	Amplitude	\\
(K)		&	(K)	&	(pW)	&	(pW)		\\
\hline
\multicolumn{4}{c}{Median Sky}	\\
\hline
 2.725  &   2.725  &  585  &   15     \\
 2.730  &   2.730  &  587  &   13     \\
 2.720  &   2.720  &  583  &   18     \\
 2.000  &  20.000  &  695  &   94     \\
 2.500  &   2.500  &  503  &   98     \\
 3.000  &   3.000  &  716  &  116     \\
\hline
\multicolumn{4}{c}{Galactic Plane}	\\
\hline
 2.725  &   2.725  &  1400  &  829     \\
 2.730  &   2.730  &  1400  &  827     \\
 2.720  &   2.720  &  1400  &  831     \\
 2.000  &  20.000  &  1510  &  719     \\
 2.500  &   2.500  &  1320  &  911     \\
 3.000  &   3.000  &  1530  &  698     \\
\hline
\multicolumn{4}{c}{Galactic Center}	\\
\hline
 2.725  &   2.725  &  4920  &  4350	\\
 2.730  &   2.730  &  4920  &  4350	\\
 2.720  &   2.720  &  4920  &  4350	\\
 2.000  &  20.000  &  5030  &  4240	\\
 2.500  &   2.500  &  4840  &  4430	\\
 3.000  &   3.000  &  5050  &  4220	\\
\hline
\end{tabular}
\end{center}
}
\end{table}

Table 1 compares the total (unmodulated) power
at the detector
to the largest (modulated) fringe amplitude
for a selection of calibrator temperatures
at three representative sky regions:
the median sky brightness,
the median Galactic plane ($|b| < 1\deg$),
and the brightest sky (Galactic center).
The principal science objectives 
observe the high-latitude sky
with the entire calibrator
within 5 mK of the CMB monopole.
As the calibrator temperature changes,
the peak fringe amplitude varies by 30\%
while the total power varies by 0.7\%.
Photon noise at the detectors
produces noise equivalent power 
2--3~$\times 10^{-16}$~W~Hz$^{-1/2}$
depending on the
high-frequency rolloff of emission from dust and zodiacal light.
Assuming detector sampling
at 256 Hz,
each time-ordered sample has white noise
with amplitude 5--7$\times 10^{-15}$~W.
Cosmological signals of interest
(Fig.~\ref{fg_figure})
are small compared to the sampled noise
and require integration over a large number of samples.
Assuming that the telemetry digitization
is set so that the noise exercises 2-3 bits,
the fringe amplitudes from the median sky
require 12--14 bits of dynamic range.
Brighter emission from the Galactic plane
or Galactic center
require additional bits
or lower post-detection gain.
Since the photon noise
is also significantly higher for these bright regions,
a post-detection gain reduction
can be employed 
to fully sample the fringe pattern from these pixels
while remaining within 16 bit digitization.

Note that the sampled fringe patterns 
simplify data compression.
Fringe amplitudes larger than 5 bits
occur only near zero phase delay
or at large sky/calibrator differences.
Roughly 95\% of the observations during flight
will sample fringe amplitudes below 5 bits:
the remaining 11 bits are identically zero
to allow a compression factor of order 70\%.
The PIXIE data rate will thus be close to 6~kbps.

\begin{figure}[t]
\centerline{
\includegraphics[height=3.0in]{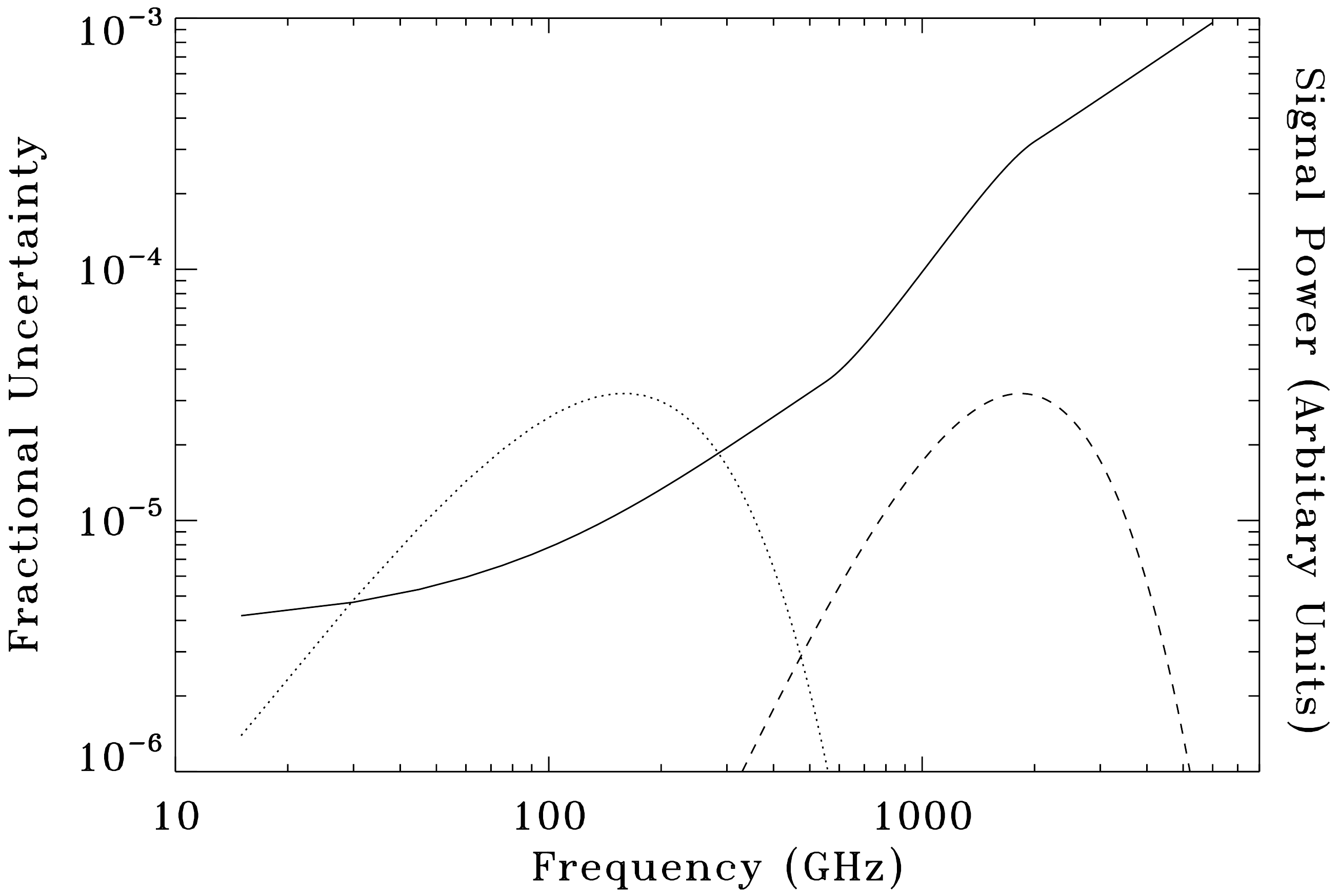}}
\caption{
The fractional uncertainty in the PIXIE calibration  
varies with frequency.
The solid line shows the combined fractional uncertainty
in the calibration.
The dotted and dashed lines
show the spectral shape
for CMB anisotropy
and the combined foreground emission,
respectively.
In-flight calibration is expected to be accurate
to a few parts in $10^6$
at frequencies dominated by the CMB,
and a few parts in $10^4$
at higher frequencies
dominated by the diffuse dust foreground.
}
\label{combined_cal_error}
\end{figure}

\section{Discussion}

A blackbody calibrator has three major performance requirements.
It must be sufficiently black and cover enough of the beam
to reduce reflections and leakage to negligible levels.
It must be sufficiently isothermal 
to reduce residual thermal gradients
to negligible levels.
The mean temperature must be determined with sufficient accuracy
that temperature errors
do not propagate to the final results.
With these three conditions,
the calibrator can be treated as a single source
whose emission is completely determined by the temperature.

The PIXIE calibrator meets these conditions.
The calibrator can be deployed to completely block
either of the two exit apertures;
the estimated leakage of -70~dB produces 
a calibration error less than $10^{-8}$.
Both the calibrator and the instrument optics
are maintained within 5~mK of CMB monopole temperature
to minimize systematic errors in the calibration.
Reflections at levels -65~dB 
terminate within the instrument blackbody cavity
to create calibration error less than $10^{-8}$.
Near-isothermal operation minimizes heat flow
which could source thermal gradients;
the predicted gradient during normal operation
is less than 1 nK 
and would be undetectably small.
Complementary data allow in-flight calibration
of both the temperature and frequency scales
to higher precision and accuracy than possible
during pre-flight testing.
Observations of bright Galactic lines
determine the frequency scale to 200 Hz accuracy,
well below the $\sim$15~GHz width of the synthesized frequency channels.
Measurements of the Wien displacement
in the peak of the blackbody calibrator spectra
fix the absolute thermometry to 10~$\mu$K accuracy
over the full range of commanded temperatures.

The PIXIE calibration is peformed {\it in situ}
using the same data as the sky measurements:
there are not separate calibration vs sky data sets
and thus no systematic differences between calibration and sky data.
Changing the calibrator temperature by a few mK
provides an absolute reference signal
at signal to noise ratio above $10^3$
on time scales of a few seconds
and $10^7$ averaged over the full mission.
Figure \ref{combined_cal_error}
shows the frequency dependence of
the PIXIE calibration uncertainty.
At frequencies below the CMB Wien cutoff,
the calibration is expected to be accurate
to a few parts in $10^6$,
limited by the 10~$\mu$K uncertainty
in the absolute temperature of the calibrator.
At higher frequencies
where the diffuse dust foreground dominates,
thermal gradients
within the calibrator
produced when the single cavity is heated to 20~K
limit the accuracy to a few parts in $10^4$.
Other effects are negligible.
The expected calibration accuracy is well under the requirements
to detect the predicted $y$ and $\mu$ spectral distortions,
which are the most stringent requirements of the PIXIE science goals.

\clearpage

%

\begin{thebibliography}{10}

\bibitem{lyth/riotto:1999}
D.~H.~D.~H. {Lyth} and A.~A. {Riotto}, \emph{{Particle physics models of
  inflation and the cosmological density perturbation}},
  \href{https://doi.org/10.1016/S0370-1573(98)00128-8}{\emph{Physics Reports}
  {\bfseries 314} (1999) 1}
  [\href{https://arxiv.org/abs/hep-ph/9807278}{{\ttfamily hep-ph/9807278}}].

\bibitem{krauss/wilczek:2014a}
L.~M. {Krauss} and F.~{Wilczek}, \emph{{From B-modes to quantum gravity and
  unification of forces}},
  \href{https://doi.org/10.1142/S0218271814410016}{\emph{International Journal
  of Modern Physics D} {\bfseries 23} (2014) 1441001}
  [\href{https://arxiv.org/abs/1404.0634}{{\ttfamily 1404.0634}}].

\bibitem{krauss/wilczek:2014b}
L.~M. {Krauss} and F.~{Wilczek}, \emph{{Using cosmology to establish the
  quantization of gravity}},
  \href{https://doi.org/10.1103/PhysRevD.89.047501}{\emph{\prd} {\bfseries 89}
  (2014) 047501} [\href{https://arxiv.org/abs/1309.5343}{{\ttfamily
  1309.5343}}].

\bibitem{distortion_wp_2019}
J.~{Chluba}, A.~{Kogut}, S.~P. {Patil}, M.~H. {Abitbol}, N.~{Aghanim},
  Y.~{Ali-Ha{\i}{\ensuremath{\ddot{}}}moud} et~al., \emph{{Spectral Distortions
  of the CMB as a Probe of Inflation, Recombination, Structure Formation and
  Particle Physics}}, {\emph{Bulletin of the American Astronomical Society}
  {\bfseries 51} (2019) 184}
  [\href{https://arxiv.org/abs/1903.04218}{{\ttfamily 1903.04218}}].

\bibitem{bicep2xkeck_2016}
{BICEP2 Collaboration} and {Keck Array Collaboration}, \emph{{Improved
  Constraints on Cosmology and Foregrounds from BICEP2 and Keck Array Cosmic
  Microwave Background Data with Inclusion of 95 GHz Band}},
  \href{https://doi.org/10.1103/PhysRevLett.116.031302}{\emph{Physical Review
  Letters} {\bfseries 116} (2016) 031302}
  [\href{https://arxiv.org/abs/1510.09217}{{\ttfamily 1510.09217}}].

\bibitem{zeldovich/sunyaev:1969}
Y.~B. {Zeldovich} and R.~A. {Sunyaev}, \emph{{The Interaction of Matter and
  Radiation in a Hot-Model Universe}},
  \href{https://doi.org/10.1007/BF00661821}{\emph{Astrophysics and Space
  Science} {\bfseries 4} (1969) 301}.

\bibitem{sunyaev/zeldovich:1970}
R.~A. {Sunyaev} and Y.~B. {Zeldovich}, \emph{{The interaction of matter and
  radiation in the hot model of the Universe, II}},
  \href{https://doi.org/10.1007/BF00653472}{\emph{Astrophysics and Space
  Science} {\bfseries 7} (1970) 20}.

\bibitem{fixsen/etal:1996}
D.~J. {Fixsen}, E.~S. {Cheng}, J.~M. {Gales}, J.~C. {Mather}, R.~A. {Shafer}
  and E.~L. {Wright}, \emph{{The Cosmic Microwave Background Spectrum from the
  Full COBE FIRAS Data Set}}, \href{https://doi.org/10.1086/178173}{\emph{\apj}
  {\bfseries 473} (1996) 576}
  [\href{https://arxiv.org/abs/astro-ph/9605054}{{\ttfamily
  astro-ph/9605054}}].

\bibitem{kogut_apc_2019}
A.~{Kogut}, M.~H. {Abitbol}, J.~{Chluba}, J.~{Delabrouille}, D.~{Fixsen}, J.~C.
  {Hill} et~al., \emph{{CMB Spectral Distortions: Status and Prospects}},  in
  \emph{Bulletin of the American Astronomical Society}, vol.~51, p.~113, Sept.,
  2019, \href{https://arxiv.org/abs/1907.13195}{{\ttfamily 1907.13195}}.

\bibitem{firas_cal_1994}
D.~J. {Fixsen}, E.~S. {Cheng}, D.~A. {Cottingham}, J.~{Eplee}, R.~E.,
  T.~{Hewagama}, R.~B. {Isaacman} et~al., \emph{{Calibration of the COBE FIRAS
  Instrument}}, \href{https://doi.org/10.1086/173577}{\emph{\apj} {\bfseries
  420} (1994) 457}.

\bibitem{dmr_cal_1992}
C.~L. {Bennett}, G.~F. {Smoot}, M.~{Janssen}, S.~{Gulkis}, A.~{Kogut},
  G.~{Hinshaw} et~al., \emph{{COBE Differential Microwave Radiometers:
  Calibration Techniques}}, \href{https://doi.org/10.1086/171363}{\emph{\apj}
  {\bfseries 391} (1992) 466}.

\bibitem{wmap_planets_2011}
J.~L. {Weiland}, N.~{Odegard}, R.~S. {Hill}, E.~{Wollack}, G.~{Hinshaw}, M.~R.
  {Greason} et~al., \emph{{Seven-year Wilkinson Microwave Anisotropy Probe
  (WMAP) Observations: Planets and Celestial Calibration Sources}},
  \href{https://doi.org/10.1088/0067-0049/192/2/19}{\emph{\apjs} {\bfseries
  192} (2011) 19} [\href{https://arxiv.org/abs/1001.4731}{{\ttfamily
  1001.4731}}].

\bibitem{planck_hfi_cal_2014}
{Planck Collaboration}, P.~A.~R. {Ade}, N.~{Aghanim}, C.~{Armitage-Caplan},
  M.~{Arnaud}, M.~{Ashdown} et~al., \emph{{Planck 2013 results. VIII. HFI
  photometric calibration and mapmaking}},
  \href{https://doi.org/10.1051/0004-6361/201321538}{\emph{\aap} {\bfseries
  571} (2014) A8} [\href{https://arxiv.org/abs/1303.5069}{{\ttfamily
  1303.5069}}].

\bibitem{wmap_cal_2003}
G.~{Hinshaw}, C.~{Barnes}, C.~L. {Bennett}, M.~R. {Greason}, M.~{Halpern},
  R.~S. {Hill} et~al., \emph{{First-Year Wilkinson Microwave Anisotropy Probe
  (WMAP) Observations: Data Processing Methods and Systematic Error Limits}},
  \href{https://doi.org/10.1086/377222}{\emph{\apjs} {\bfseries 148} (2003) 63}
  [\href{https://arxiv.org/abs/astro-ph/0302222}{{\ttfamily
  astro-ph/0302222}}].

\bibitem{planck_lfi_cal_2013}
{Planck Collaboration}, N.~{Aghanim}, C.~{Armitage-Caplan}, M.~{Arnaud},
  M.~{Ashdown}, F.~{Atrio-Barandela} et~al., \emph{{Planck 2013 results. V. LFI
  calibration}}, \href{https://doi.org/10.1051/0004-6361/201321527}{\emph{\aap}
  {\bfseries 571} (2014) A5} [\href{https://arxiv.org/abs/1303.5066}{{\ttfamily
  1303.5066}}].

\bibitem{kogut/etal:2011}
A.~{Kogut}, D.~J. {Fixsen}, D.~T. {Chuss}, J.~{Dotson}, E.~{Dwek}, M.~{Halpern}
  et~al., \emph{{The Primordial Inflation Explorer (PIXIE): a nulling
  polarimeter for cosmic microwave background observations}},
  \href{https://doi.org/10.1088/1475-7516/2011/07/025}{\emph{Journal of
  Cosmology and Astroparticle Physics} {\bfseries 7} (2011) 25}
  [\href{https://arxiv.org/abs/1105.2044}{{\ttfamily 1105.2044}}].

\bibitem{boggess/etal:1992}
N.~W. {Boggess}, J.~C. {Mather}, R.~{Weiss}, C.~L. {Bennett}, E.~S. {Cheng},
  E.~{Dwek} et~al., \emph{{The COBE Mission: Its Design and Performance Two
  Years after Launch}}, \href{https://doi.org/10.1086/171797}{\emph{\apj}
  {\bfseries 397} (1992) 420}.

\bibitem{pan/etal:2019}
Z.~{Pan}, M.~{Liu}, R.~{Basu Thakur}, B.~A. {Benson}, D.~J. {Fixsen},
  H.~{Goksu} et~al., \emph{{Compact millimeter-wavelength Fourier-transform
  spectrometer}}, \href{https://doi.org/10.1364/AO.58.006257}{\emph{Applied
  Optics} {\bfseries 58} (2019) 6257}
  [\href{https://arxiv.org/abs/1905.07399}{{\ttfamily 1905.07399}}].

\bibitem{arcade_cal_2006}
D.~J. {Fixsen}, E.~J. {Wollack}, A.~{Kogut}, M.~{Limon}, P.~{Mirel},
  J.~{Singal} et~al., \emph{{Compact radiometric microwave calibrator}},
  \href{https://doi.org/10.1063/1.2209960}{\emph{Review of Scientific
  Instruments} {\bfseries 77} (2006) 064905}.

\bibitem{steelcast_2008}
E.~J. {Wollack}, D.~J. {Fixsen}, R.~{Henry}, A.~{Kogut}, M.~{Limon} and
  P.~{Mirel}, \emph{{Electromagnetic and Thermal Properties of a Conductively
  Loaded Epoxy}},
  \href{https://doi.org/10.1007/s10762-007-9299-4}{\emph{International Journal
  of Infrared and Millimeter Waves} {\bfseries 29} (2008) 51}.

\bibitem{nagler/etal:2015}
P.~C. {Nagler}, D.~J. {Fixsen}, A.~{Kogut} and G.~S. {Tucker},
  \emph{{Systematic Effects in Polarizing Fourier Transform Spectrometers for
  Cosmic Microwave Background Observations}},
  \href{https://doi.org/10.1088/0067-0049/221/1/21}{\emph{\apjs} {\bfseries
  221} (2015) 21} [\href{https://arxiv.org/abs/1510.08089}{{\ttfamily
  1510.08089}}].

\bibitem{fixsen/etal:2011}
D.~J. {Fixsen}, A.~{Kogut}, S.~{Levin}, M.~{Limon}, P.~{Lubin}, P.~{Mirel}
  et~al., \emph{{ARCADE 2 Measurement of the Absolute Sky Brightness at 3-90
  GHz}}, \href{https://doi.org/10.1088/0004-637X/734/1/5}{\emph{\apj}
  {\bfseries 734} (2011) 5} [\href{https://arxiv.org/abs/0901.0555}{{\ttfamily
  0901.0555}}].

\bibitem{kogut/etal:2019}
A.~J. {Kogut} and D.~J. {Fixsen}, \emph{{Systematic error cancellation for a
  four-port interferometric polarimeter}},
  \href{https://doi.org/10.1117/1.JATIS.5.2.024008}{\emph{Journal of
  Astronomical Telescopes, Instruments, and Systems} {\bfseries 5} (2019)
  024008} [\href{https://arxiv.org/abs/1908.00558}{{\ttfamily 1908.00558}}].

\bibitem{fixsen/etal:2009}
D.~J. {Fixsen}, \emph{{The Temperature of the Cosmic Microwave Background}},
  \href{https://doi.org/10.1088/0004-637X/707/2/916}{\emph{\apj} {\bfseries
  707} (2009) 916} [\href{https://arxiv.org/abs/0911.1955}{{\ttfamily
  0911.1955}}].

\bibitem{herschel_2013}
J.~L. {Pineda}, W.~D. {Langer}, T.~{Velusamy} and P.~F. {Goldsmith}, \emph{{A
  Herschel [C ii] Galactic plane survey. I. The global distribution of ISM gas
  components}}, \href{https://doi.org/10.1051/0004-6361/201321188}{\emph{\aap}
  {\bfseries 554} (2013) A103}
  [\href{https://arxiv.org/abs/1304.7770}{{\ttfamily 1304.7770}}].

\bibitem{planck_fg_2018}
{Planck Collaboration}, Y.~{Akrami}, M.~{Ashdown}, J.~{Aumont},
  C.~{Baccigalupi}, M.~{Ballardini} et~al., \emph{{Planck 2018 results. IV.
  Diffuse component separation}}, {\emph{arXiv e-prints} (2018)
  arXiv:1807.06208} [\href{https://arxiv.org/abs/1807.06208}{{\ttfamily
  1807.06208}}].

\bibitem{fixsen/etal:1998}
D.~J. {Fixsen}, E.~{Dwek}, J.~C. {Mather}, C.~L. {Bennett} and R.~A. {Shafer},
  \emph{{The Spectrum of the Extragalactic Far-Infrared Background from the
  COBE FIRAS Observations}}, \href{https://doi.org/10.1086/306383}{\emph{\apj}
  {\bfseries 508} (1998) 123}
  [\href{https://arxiv.org/abs/astro-ph/9803021}{{\ttfamily
  astro-ph/9803021}}].

\end{thebibliography}

\providecommand{\href}[2]{#2}\begingroup\raggedright\endgroup

\end{document}